\documentclass[11pt]{article}
\usepackage{snapshot}
\pdfoutput=1

\usepackage{amsmath}
\usepackage{latexsym}
\usepackage{amsfonts}
\usepackage{graphics}
\usepackage{amssymb}
\usepackage{latexsym}
\usepackage{epsfig}
\usepackage{natbib}
\usepackage[pdftex]{color}
\definecolor{mired}{rgb}{1,0,0}

\usepackage[ruled]{algorithm2e}
\usepackage{amsmath}

\newcommand{\bfx}{{\mbox{\boldmath $x$}}}

\newcommand{\beq}{\begin{equation}}
\newcommand{\eeq}{\end{equation}}

\newcommand{\beqa}{\begin{eqnarray}}
\newcommand{\eeqa}{\end{eqnarray}}

\newcommand{\bpm}{\begin{pmatrix}}
\newcommand{\epm}{\end{pmatrix}}
\newcommand{\nn}{\nonumber}

\begin{document}

\title{Communications with decode-and-forward relays in mesh networks}

\author{ \'Angel Bravo-Santos\\
Universidad Carlos III de Madrid,\\
\texttt{abravo@tsc.uc3m.es} \and 
Petar M. Djuri\'c\\
Stony Brook University\\
\texttt{djuric@ece.sunysb.edu}}

\maketitle

\begin{abstract}
  We consider mesh networks composed of groups of relaying nodes which operate in decode-and-forward mode, where each node from a group relays information to all the nodes in the next group. We study these networks in two setups, one where the nodes have complete channel state information from the nodes that transmit to them, and another when they only have the statistics of the channel. We derive recursive expressions for the probabilities of errors of the nodes and present several implementations of detectors used in these networks. We compare the mesh networks with multihop networks, the latter being formed by a set of parallel sections of multiple relaying nodes. We demonstrate with numerous simulations that there are significant improvements in performance of mesh over multihop networks in various scenarios.

\end{abstract}

%\begin{IEEEkeywords}
%decode-and-forward, mesh networks, multi-hop networks
%\end{IEEEkeywords}
%\newpage

%\ifCLASSOPTIONpeerreview
%\begin{center} \bfseries EDICS Category: SEN-COLB, SEN-DIST, SPC-DETC, WIN-CONT  \end{center}
%\fi

%\IEEEpeerreviewmaketitle

\section{Introduction}

%The topologies of wireless networks keep evolving so that they can
%support ever more demanding services in areas where they are
%used. From a classical point of view, wireless networks are formed by two entities: one is the access network formed by base stations
%communicating with subscriber stations and the other is the transport network linking the base stations. The differences between these two types of networks is disappearing and the stations, nodes in general, can be part of the transport network or can even act as subscriber stations.

An old concept in radio communications, known as relaying, has been in the center of interest of various studies in communications
\citep{ephremides02,goldsmith02,ribeiro08}. %Perhaps 
To a good extent, this interest has been driven by %the main reason is that wireless networks have been of significant research and
commercial applications and %for several years and that 
by the ability of wireless networks to exploit relaying so that they have reduced energy consumption and thereby increased lifetime. In general, relaying is used to provide for improved error performance and capacity \citep{liu09}.
%For example, it is well known that the use of links with multiple antennas over multiple-input multiple-output (MIMO) channels can considerably improve the performance of the system \citep{foschini98}.

In networks with relaying, nodes cooperate in moving information from source to destination \citep{liu09}. In the study of these  networks, the important notion of cooperative diversity was introduced in \citep{sendonaris98}, and then extended in \citep{sendoranis03} and \citep{laneman04}. A class of wireless networks that are formed by a set of parallel sections of multiple relaying nodes was thoroughly studied in the literature \citep{lin05,wang07}. They are known as multi-branch multi-hop networks, and they can reach a full diversity order \citep{wang07}. However, unlike mesh networks, these networks do not allow for full connectivity among nodes.

Optimum detection in a network demands knowledge of all its
channels. There is some work on this subject in the literature, and especially on particular network topologies. For example, in a recent paper, the problem of detection by a set of sensors and the
communication of their information were considered together
\citep{chen04}. Various fusion rules were proposed that correspond to
different scenarios in terms of what is known for deriving the
rules. The study was performed for multi-branch multi-hop topologies, but not for general mesh networks.

An analysis of the maximum likelihood (ML) detector in cooperative multibranch multi-hop
networks was presented in \citep{chen06}. A closed form expression was obtained for the bit error rate of a network with one relay. The
difficulties with ML detection in DF networks have led to several
suboptimal alternatives. In \citep{sendoranis03_2}, a variant of the
maximal-ratio combiner (MRC), the $\lambda$-MRC, was introduced to
combine signals from several branches. In order to explicitly obtain a full diversity order, the cooperation MRC (CMRC) was proposed in
\citep{wang07}. For this detector, the weight of the channel of
relay-destination was chosen to maximize the equivalent SNR for the
channel. In that paper the multibranch multi-hop topology was
considered. It is interesting to mention that a multi-node cooperative network can be viewed as a virtual MIMO system
\citep{liu09,sadek07}. In successive phases the relays combine the
received signals from previous relays and the source using the MRC
criterion. The system is not optimal and it can be adapted to a mesh
network.

Most of the above solutions assume perfect knowledge of the channel
state information (CSI). However, having such knowledge can be
 expensive. For instance, in a sensor network the battery consumption can be high if the nodes have to inform about their CSI in a variable environment. In other circumstances it could be impossible to have complete knowledge of the CSI.  For these cases it is important to develop schemes that operate with less information about the channel: instead of full information one has available only the channel statistics. There is some previous work in the literature on detection using channel statistics. In \citep{bravo06}, the average SNR is used as a weight for combining signals in a multi-branch multi-hop relay network. In \citep{lin05}, optimum ML fusion rules are derived for a joint sensor-communication problem with a multi-branch multi-hop topology.

%The offered traffic to sensor networks is usually small.
Data traffic in a sensor network is usually small. We take this statement as true, although it is not always the case, and we assume that most of the available channels can be used for cooperation and we do not put a limit to this number. This is a usual assumption in the literature, see \citep{wang07}. A consequence of this assumption is that the spectral efficiency is reduced. In the literature, there are examples of simple networks where the error probability and the spectral efficiency are improved at the same time \citep{sendoranis03_2}. However, it is not straightforward to extend these results to arbitrary networks. In this paper, we look for the objective of minimizing the transmitted power, or, what is related to it, minimizing the error probability, with no restrictions in the number of available channels.

The cooperative detection problems mentioned above were posed for multibranch and/or multi-hop networks. For them, the
optimal ML solution was abandoned in favor of tractable solutions. In this paper we present and analyze the optimal maximum a posteriori (MAP) detector for cooperative mesh networks, and we provide explicit analytical solution for a general topology. We consider optimal detection in two cases related to the channels in the network. One is when the CSI is available and the other when the channel statistics are only known. The first case was studied in part by the authors in \citep{bravo09}. The relaying in the network is DF with uncoded and symbol-by-symbol demodulators. In summary, the two main contributions in our paper are the following: (a) we derive optimal MAP detectors for the nodes in mesh networks and (b) we propose several implementations of the detectors for both, the case of known CSI and scenarios of known channel statistics. In the paper we do not consider the problems of routing and protocols that may arise, for example, in ad hoc mesh networks. How they relate to the proposed schemes here will be addressed elsewhere.

%Usually the environment and electric parameters affecting the quality of wireless networks can be optimized, for instance, the nodes sites, antenna gains or the number of hops of a link. However, in sensor networks randomly, or quasi-randomly, deployed the mentioned variables are not under the control of the network planner. Even more, channel gains can suffer random variations due to the movement of the nodes or the objects close to them. For this reason in this paper we consider that channel gains are data, and not variables that are part of the input for the optimization the network.

The paper is organized as follows. In Section \ref{sec:ps} we
formulate the problem. In Section \ref{sec:der}, we provide the
general MAP solution that holds for known CSI as well as for known
channel statistics. In Sections \ref{sec:cha_des} and \ref{sec:imple} we present detectors in the first and higher groups of nodes, respectively, and provide the specific solutions for the above two scenarios. Simulation results that demonstrate the performance of the mesh networks and how they compare with the multibranch and multi-hop networks are shown in Section \ref{sec:sim}. In the last Section \ref{sec:con}, we have some concluding remarks about our findings.

\section{Problem statement}
\label{sec:ps}

\begin{figure}[thb]
\centerline
    {
\epsfxsize=\linewidth \epsffile{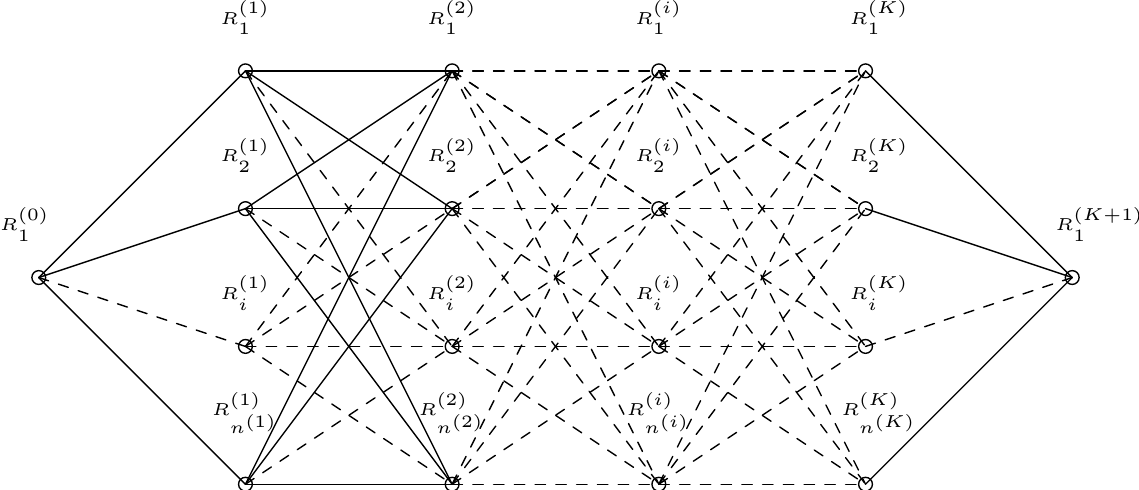}
    }
%\vspace*{-0.4cm}
\caption{A general mesh network with $K$ groups of relaying nodes.}
%\vspace*{0.5cm}
\label{fig:mesh}
\end{figure}

We observe a mesh network where its nodes are grouped into relay
groups and where the notation $R_i^{(k)}$ signifies the $i$th node of the $k$th group. We denote the source by $R_1^{(0)}$ and the
destination by $R_1^{(K+1)}$, the superscript $K+1$ implying that a
message is relayed $K$ times on its way from the source to the
destination. For example, $R_4^{(3)}$ refers to the fourth node from
the third group of nodes. We use the symbol $n^{(k)}$ to denote the
total number of relay nodes in group $k$.  In Fig. \ref{fig:mesh} we show a drawing of a general mesh network of our interest.

\begin{figure}[thb]
\centerline
    {
\epsfxsize=\linewidth \epsffile{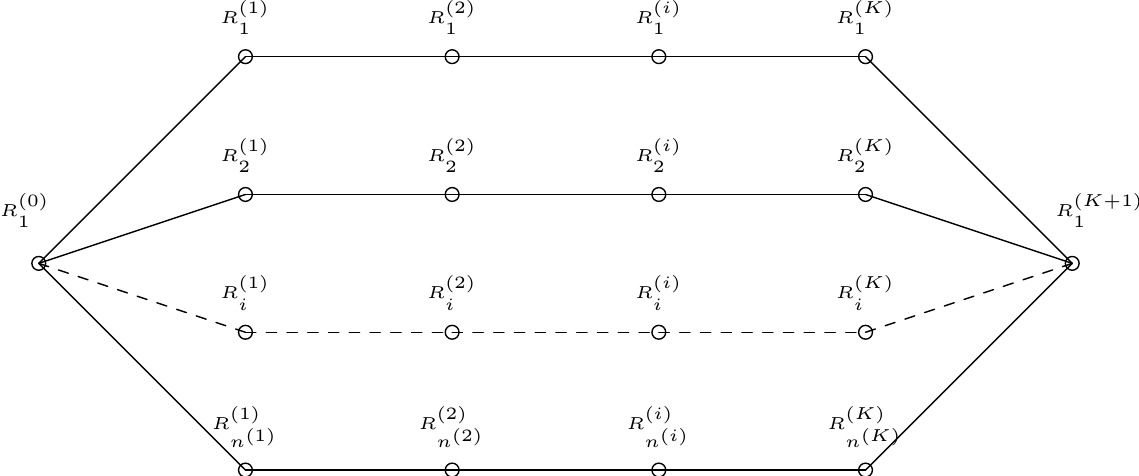}
    }
%\vspace*{-0.4cm}
\caption{A general multi-branch multi-hop  network with $K$ groups of relaying nodes.}
%\vspace*{0.5cm}
\label{fig:mhop}
\end{figure}

In this paper, for comparison purposes, we also work with multi-branch multihop networks. %We work with them in this paper for comparisons with mesh networks and when the mesh network has equal number of nodes.
Multi-branch multihop networks can be obtained from mesh networks when we remove connections between nodes so that each node in a group is connected to only one node from the previous group, as in Fig. \ref{fig:mhop}. Clearly, a wireless network formed by a set of interconnected nodes can operate as a mesh network or as a multi-branch multihop network. In mesh networks, a node receives and processes information from more than one node from a previous relay group, whereas in a multihop network, it does only from one node.

We consider binary modulations only (the extension to other
modulations is straightforward), and we assume phase coherent
reception. The received signal by node $R_j^{(k)}$ and transmitted by node $R_i^{(k-1)}$, where $k\geq 1$, is denoted by
\beqa
\label{eq:model}
y_{ij}^{(k)}&=&h_{ij}^{(k)}x_i^{(k-1)}+w_{ij}^{(k)},
\eeqa
where
$x_i^{(k-1)}\in \{-1, 1\}$ is the transmitted symbol, $h_{ij}^{(k)}$
is the channel fading between $R_i^{(k-1)}$ and $R_j^{(k)}$, and
$w_{ij}^{(k)}$ is the channel noise.  In this formulation,
$h_{ij}^{(k)}>0$ is the real-valued fading envelope of the channel. We denote the transmitted symbol by the source with $x$ and not $x_1^{(0)}$ so that we avoid superfluous notation.  The noise is modeled as a zero mean Gaussian random variable with variance $\sigma^2$, i.e., $w_{ij}^{(k)}\sim {\cal N}(0, \sigma^2)$, and it is considered identically distributed in all the channels.

We assume a simple protocol where every node has access to an
orthogonal channel. This can be implemented with time division,
frequency division or hybrid multiplexing. %, where each node has
%an assigned time and/or frequency channel.
This assumption could be considered unrealistic, but with a little traffic offered in a network, as is frequently the case in sensor networks, most of the available channels can be used for cooperation
with the node source of information and the time/frequency channels can be reused by distant nodes. We also point out that in a real system the number of available channels is not necessarily small. For example, the Zigbee standard for sensor networks \citep{zigbee10} envisages a hybrid access for its next release in order to increase the number of channels and, thus, reduce message collisions.

We distinguish two scenarios:
\begin{enumerate}
\item all the channels in the network are completely known (CSI is available to all the nodes), and
\item only the statistics of the channels are known, where it is assumed that the channels follow the Rayleigh distribution, i.e., for the fading of a channel $h$ we write
 \beqa
 \label{rayleigh}
h&\sim& \frac{{2} h}{\sigma_{h}^2}e^{-\frac{h^2}{\sigma_{h}^2}},\;\;h\geq 0.
\eeqa
\end{enumerate}
The signal-to-noise ratio (SNR) in the former case is defined by
\beqa
\label{snr}
\gamma&=&\frac{h^2}{\sigma^2},
\eeqa
and in the latter, by
\beqa
\gamma&=&\frac{\sigma_h^2}{\sigma^2}.
\eeqa
Obviously, we can have a combination of the two scenarios, but we will not consider it here.

Given the available information, the objective is to determine the optimal decisions at each node and study the performance of the systems under the two scenarios.

\section{General solution}
\label{sec:der}

In this section we derive the decision rule of the nodes under the
assumptions from the previous section.
%We assume that each node in the network has the information about the probabilities of correct and incorrect decisions of the nodes from the previous group.
We denote the probability of correct decision of node $i$ in the $k$th group by
$P^{(k)}_{i}=P(x_i^{(k)}=x|x)$ and the probability of error by
$\overline{P}^{(k-1)}_{i}=P(x_i^{(k-1)}=-x|x)$. In the sequel, we
assume $P(x=1)=P(x=-1)$.

Clearly, the nodes $R_j^{(1)}$, $j= 1, 2, \cdots, n^{(1)}$  make their decisions according to
%\beqa
%L_j^{(1)}&=&\frac{P\left(x=1|y_{1j}^{(1)}\right)}{P\left(x=-1|y_{1j}^{(1)}\right)}\nonumber\\
%&=&\frac{f(y_{1j}^{(1)}|x=1)}{f(y_{1j}^{(1)}|x=-1)},
%\eeqa
\beq
L_j^{(1)}=\frac{P\left(x=1|y_{1j}^{(1)}\right)}{P\left(x=-1|y_{1j}^{(1)}\right)}
=\frac{f(y_{1j}^{(1)}|x=1)}{f(y_{1j}^{(1)}|x=-1)},
\eeq
where $j = 1, 2, \cdots, n^{(1)}$, and $f(y_{1j}^{(1)}|x=s)$ is the likelihood of $x=s$, with $s$ being $1$ or $- 1$.

The following groups of nodes  will receive in general more than one
signal and a decision is made by using all of them. Let the signals
received by node $R_j^{(k)}$ be denoted by the vector ${\bf
  y}_j^{(k)}=[y_{1j}^{(k)}, y_{2j}^{(k)}, \cdots, y_{n^{(k-1)}j}^{(k)}
]^\top$, and let the decisions of the nodes in the previous group, be given by ${\bf x}^{(k-1)}=[x_{1}^{(k-1)}, x_{2}^{(k-1)}, \cdots,
x_{n^{(k-1)}}^{(k-1)} ]^\top$.  Then we have
%\beqa
\begin{align}
  L_j^{(k)}&=\frac{P\left(x=1|{\bf y}_{j}^{(k)}\right)}{P\left(x=-1|{\bf y}_{j}^{(k)}\right)}\notag=\frac{f({\bf y}_{j}^{(k)}|x=1)}{f({\bf y}_{j}^{(k)}|x=-1)}\notag
\end{align}
    \begin{align}
&=\frac{\sum_{{\bf x}^{(k-1)}}f({\bf y}_{j}^{(k)}|{\bf
    x}^{(k-1)})P({\bf x}^{(k-1)}|x=1)}{\sum_{{\bf x}^{(k-1)}}f({\bf
    y}_{j}^{(k)}|{\bf x}^{(k-1)})P({\bf
    x}^{(k-1)}|x=-1)}\notag\\
&=\frac{\sum_{{\bf x}^{(k-1)}}P({\bf x}^{(k-1)}|x=1)\prod_{i=1}^{n^{(k-1)}}f({y}_{ij}^{(k)}|{
    x_i}^{(k-1)})}{\sum_{{\bf x}^{(k-1)}}P({\bf
    x}^{(k-1)}|x=-1)
    \prod_{i=1}^{n^{(k-1)}}f({y}_{ij}^{(k)}|{
    x_i}^{(k-1)})}.
\label{express} %\eeqa
\end{align}

The decision rule at node $R_j^{(k)}$ is
\beqa
\label{eq:dec_rul}
  x_j^{(k)}& =& \operatorname{sgn}\left(\log \big(L_j^{(k)}\big)\right).
\eeqa
For $k>2$ and general mesh networks, the decision rule \eqref{eq:dec_rul} requires that a node must know the states or statistics of the channels {\em and} the joint probability mass function (pmf) of correct decision of its relaying nodes, $P({\bf x}^{(k-1)}|x=1)$ and $P({\bf x}^{(k-1)}|x=-1)$. The former requirement in most studies is assumed satisfied. The second requirement is much more restrictive, and in the sequel we propose ways of dealing with it.

%However, there are two practical problems related with the detection: a) the number of values of $P({\bf x}^{(k-1)}|x=1)$ and $P({\bf x}^{(k-1)}|x=-1)$  is exponential en the number of nodes per group, b) in order to evaluate the probability of correct decision of all the nodes of a group, $k$ for instance, the information about all the channels associated with that group must be available at some point, for instance, a particular node of the group. This is difficult, if possible, in a real network. But it is not a problem in a simulation environment where a mesh network can be programmed for obtaining results about its best case performance.

%%%%%%%%%%%%%%%%%%%%%%%%%%%%
%OJO
We note that in \eqref{express}, we have likelihoods of the form $f({\bf y}_{j}^{(k)}|x)$. For them we have the following claim:

{\em Claim 1}: For the likelihood $f({\bf y}_{j}^{(k)}|x)$, $k=1,2$ in mesh networks, we can write
\beqa
\label{claim}
f\left({\bf y}_{j}^{(k)}|x\right)&=&\prod_{i=1}^{n^{(k)}} f\left({y}_{ij}^{(k)}|x\right).
\eeqa

Proof: The proof follows directly from the conditional independence  of the channels and the decision independence of the nodes in the first group.
\medskip

From Claim 1, for the likelihood ratio  in nodes of group 1 and 2, $L_{ij}^{(k)}$, $k=1, 2$; $j= 1, 2, \cdots, n^{(k)}$, we can write
\beqa
L_j^{(k)}&=&\prod_{i=1}^{n^{(k-1)}} L_{ij}^{(k)}%\nn\\
=\prod_{i=1}^{n^{(k-1)}}\frac{f({y}_{ij}^{(k)}|x=1)}{f({y}_{ij}^{(k)}|x=-1)}\nn\\
&=&\prod_{i=1}^{n^{(k-1)}}\frac{f({y}_{ij}^{(k)}|x_i^{(k-1)}=1)
{P}_i^{(k-1)}+f({y}_{ij}^{(k)}|x_i^{(k-1)}=-1)\overline{P}_i^{(k-1)}}
{f({y}_{ij}^{(k)}|x_i^{(k-1)}=-1){P}_i^{(k-1)}+f({y}_{ij}^{(k)}|x_i^{(k-1)}=1)
\overline{P}_i^{(k-1)}}\label{genexp},
\eeqa
where
%\beqa
\begin{align}
P_i^{(k-1)}&=P(x_i^{(k-1)}=1|x=1) %\notag\\&
=P(x_i^{(k-1)}=-1|x=-1)\label{eq:sim_1}
%\eeqa
\end{align}
is the probability of correct decision, and
\beq
\label{eq:sim_2}
\overline{P}_i^{(k-1)}=1-{P}_i^{(k-1)}
\eeq
is the probability of error, with $P^{(0)}=1$.

The likelihood ratio in \eqref{genexp}, in general, does not have the same simple form for $k>2$  because of the dependence among the decisions of the relaying nodes. However, if we approximate
\beqa
\label{eq:assumption}
P\left({\bf x}^{(k-1)}|x\right)&\approx&\prod_{i=1}^{n^{(k-1)}}P\left(x_i^{(k-1)}|x\right),
\eeqa
%If, we decide to ignore the interdependence of the decisions of the relaying group of nodes and use only their marginal probabilities  the  is useful for
we will obtain suboptimal detectors that use \eqref{genexp}, as is shown in Section \ref{sec:imple}. We note that with the assumption \eqref{eq:assumption}, we only need to know the marginal probabilities of correct decisions of the relaying nodes.

A symmetry property, similar to \eqref{eq:sim_1}  and
\eqref{eq:sim_2}, can be stated for $P({\bf x}^{(k-1)}|x=1)$ and
$P({\bf x}^{(k-1)}|x=-1)$.  %If we define $\bar{\mathbf{x}}^{(k)}$ as
                           %the vector with the same components as
                           %$\mathbf{x}^{(k)}$ but with a changed
                           %sign,
We make the following claim:

{\em Claim 2}: If the channel likelihoods satisfy
$f\big(y_{ij}^{(k)}|x_i^{(k-1)}\big)=f\big(-y_{ij}^{(k)}|-x_i^{(k-1)}\big)$
and
$f\big(-y_{ij}^{(k)}|x_i^{(k-1)}\big)=f\big(y_{ij}^{(k)}|-x_i^{(k-1)}\big)$
for all $k$,
we have
\begin{equation}
  \label{eq:clm2}
  P\left({\mathbf{x}}^{(k)}|x=1\right)=P\left(-{\mathbf{x}}^{(k)}|x=-1\right).
\end{equation}

Proof: See Appendix 1.
\medskip

We note that if the conditions for Claim 2 are satisfied, for optimal processing of the received signals all the information about the joint pmf of correct decisions needed by a node in the $k$th group is in $P(\mathbf{x}^{(k-1)}|x=1)$. Then, for the likelihood ratio \eqref{express} we can formally write
\begin{equation}
  \label{eq:lrt_df}
   L_j^{(k)}=\frac{\sum_{{\bf x}^{(k-1)}}P({\bf x}^{(k-1)}|x=1)\prod_{i=1}^{n^{(k-1)}}f({y}_{ij}^{(k)}|{
    x_i}^{(k-1)})}{\sum_{{{\bf x}}^{(k-1)}}P({-{\mathbf{x}}}^{(k-1)}|x=1)
    \prod_{i=1}^{n^{(k-1)}}f({y}_{ij}^{(k)}|{
    x_i}^{(k-1)})},
\end{equation} %

where all the probabilities are conditioned on $x=1$.

\section{Detectors in the first group of nodes}
\label{sec:cha_des}

In Section \ref{sec:ps}, we described two scenarios, one where the necessary CSI is available to the nodes, and another where only statistics of the channels are known. Here we describe the detectors of the nodes in the first group and we present their performances.

\subsection{Completely known channels}
\label{sec:com_csi}

In the case of known channels and based on the assumptions from Section \ref{sec:ps}, we can express the likelihood as follows:
\begin{equation}
  \label{eq:norm_1}
  f\left(y_{1j}^{(1)}|h_{1j}^{(1)},x\right)=\frac{1}{\sqrt{2\pi\sigma^2}}
  e^{-\frac{\left(y_{1j}^{(1)}-h_{1j}^{(1)}x\right)^2}{2\sigma^2}}.
\end{equation}
%The distributions $f\left(\mathbf{y}_{j}^{(k)}|x\right)$ are mixture Gaussians given by
%\begin{equation}
%  \label{eq:pynorm}
%  P(\mathbf{y}_j^{(k)} |x)=
%  \sum_{\mathbf{x}^{(k-1)}} P(\mathbf{x}^{(k-1)}|x)
%\prod_{i=1}^{n^{(k-1)}}
%          {\cal N}({h}_{ij}^{(k)} {x}_i^{(k-1)},\sigma^2).
%\end{equation}
The decision rule of node $R_j^{(1)}$ is based on
\beqa
L_j^{(1)}&=&\frac{p(y_{1j}^{(1)}|x=1)}{p(y_{1j}^{(1)}|x=-1)},
\eeqa
which simplifies to
\beqa
\label{rule-1}
x_j^{(1)}&=&\operatorname{sgn}\left(y_{1j}^{(1)}\right).
\eeqa
%\beqa
%\label{rule-1}
%x_j^{(1)}&=&\left\{\begin{array}{rl}
%1, & y_{1j}^{(1)}\geq 0\\
%-1, &y_{1j}^{(1)}<0
%\end{array}\right..
%\eeqa
For the nodes in this group, we can easily find the probability of error, and it is given by
\beqa
\overline{P}_j^{(1)}&=&Q\left(\sqrt{\gamma_{1j}^{(1)}}\right)\label{prob_c},
\eeqa
with $Q(z)=1-\Phi(z)$, where $\Phi(z)=\int_{-\infty}^z \frac{1}{\sqrt{2\pi}}e^{-\frac{t^2}{2}}dt$, and $\gamma_{1j}^{(1)}$ is the SNR of the channel that links the transmitter and the $j$th node in group one (see \eqref{snr}).

%The decision rules for nodes in groups 2 and following are described in Section \ref{sec:imple}

\subsection{Channels with known statistics}
\label{sec:kno_stat}

When the channels are not known, and instead the nodes only have available the channel statistics, we proceed as follows \citep{niu06}. We write for the likelihoods
\beqa
f\left(y_{1j}^{(1)}|x\right)&=&\int_0^\infty f\left(y_{1j}^{(1)}|h_{1j}^{(1)}, x \right) f(h_{1j}^{(1)}) dh_{1j}^{(1)},
\eeqa
where $f(h_{1j}^{(k)})$ is given by \eqref{rayleigh}.

For simplicity, we rewrite the last integral without subscripts and superscripts, and we get
\beqa
f(y|x)&=&\int_0^\infty  \frac{1}{\sqrt{2\pi\sigma^2}}\exp\left(-\frac{(y-hx)^2}{2\sigma^2}\right)
\frac{2h}{\sigma_h^2}\;\exp\left(-\frac{h^2}{\sigma_h^2}\right)dh\nonumber.
\eeqa
It is easy to show that $f(y|x)$ can further be expressed as
\beqa
f(y|x)\label{integral}
&=&\sqrt{\frac{2}{\pi\sigma^2\sigma_h^4}}\;\exp\left(-\frac{y^2}{2\sigma^2}\right)\int_0^\infty
h\exp\left(-\mu h^2+2\nu h\right)dh
\eeqa
where
\beq
\mu=\frac{x^2}{2\sigma^2}+\frac{1}{\sigma_h^2}, \
\nu=\frac{y\,x}{2\sigma^2}.
\eeq
We can show that the integral in the above equation can be analytically solved. We get
\beqa
\label{solution_1}
\int_0^\infty
h\exp\left(-\mu h^2+2\nu h\right)dh&=&\frac{1}{2\mu}+\frac{\nu}{\mu}\sqrt{\frac{\pi}{\mu}}\exp\left( \frac{\nu^2}{\mu}\right)\Phi\left(\nu\sqrt{\frac{2}{\mu}}\right).
\eeqa

After substituting the solution \eqref{solution_1} in \eqref{integral} and returning all the superscripts and subscripts, we obtain
\beqa
\label{eq:pdf}
f(y_{1j}^{(1)}|x)&=&\sqrt{\frac{2}{\pi}}\frac{\sigma^3a^2}{\sigma_h^2}\exp\left(-\frac{y_{1j}^{(1)^2}}
{2\sigma^2}\right)
\left(1+\sqrt{2\pi}a\,x\, y_{1j}^{(1)}\Phi\left(a\,x\, y_{1j}^{(1)}\right)
\exp\left(\frac{a^2y_{1j}^{(1)^2}}{2}\right) \right),
\eeqa
where
\beqa
a&=&\frac{\sigma_h}{\sigma\sqrt{2\sigma^2+\sigma_h^2}}\nn.
\eeqa

\begin{figure}[thb]
\centerline
    {
\epsfxsize=\linewidth \epsffile{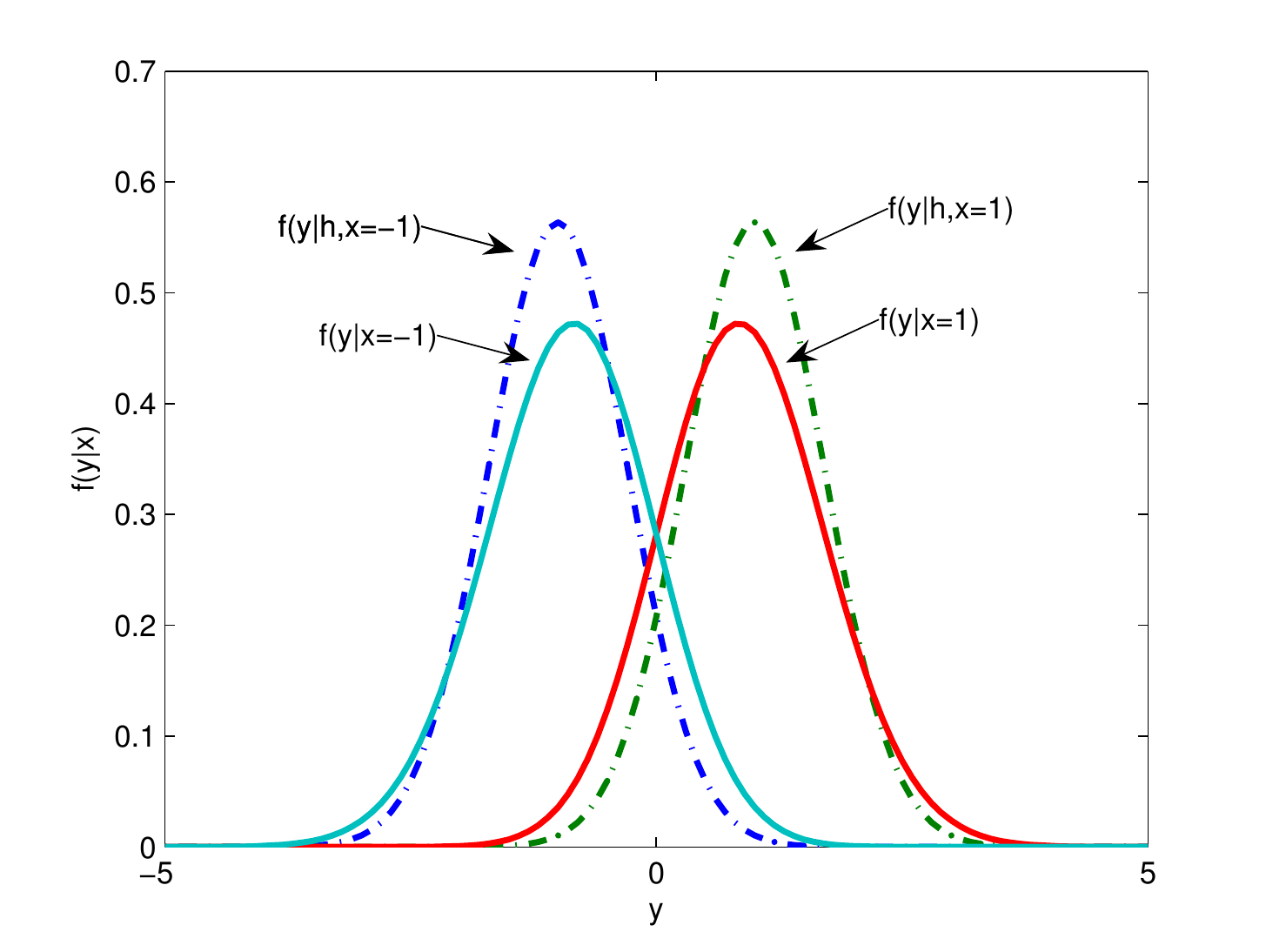}
    }
%\vspace*{-0.4cm}
\caption{The pdfs of $y$, $f(y|x)$, for $x=\pm1$ when the channels are known (curves with dash-dot lines) and when the statistics of the channels are known (solid lines) for SNR equal to 3dB. }
%\vspace*{0.5cm}
\label{fig-pdfs}
\end{figure}

In Fig. \ref{fig-pdfs}, we plotted the probability density functions
(pdfs) of $y$ given $x=\pm 1$ for SNR $=$ 3dB  for the two cases of
interest, when the channels are completely known and when their
statistics are only known. As expected, the latter case leads to
deterioration in performance because then the pdfs are shifted towards
$y=0$ and are flatter in comparison to the pdfs that correspond to known channels.

For the probability of error at node $R_j^{(1)}$, we have
\beqa
\overline{P}_j^{(1)}  %&=&P\left(l_{j}^{(1)}<0\big|x=1\right)\nn\\
%&=&P\left(L_{1j}^{(1)}<1\big|x=1\right)\nn\\
%&=&P\left(y<0\right|x=1)\label{Pe1_0},
=P\left(L_{1j}^{(1)}<1\big|x=1\right)
&=&P\left(y<0\right|x=1)\label{Pe1_0},
\eeqa
where
\beqa
y_{1j}^{(1)}&\sim& f(y_{1j}^{(1)}|x=1)\nn,
\eeqa
i.e.,
\beqa
\overline{P}_j^{(1)}&=&\int_{-\infty}^0
\sqrt{\frac{2}{\pi}}\frac{\sigma^3a^2}{\sigma_h^2}\exp\left(-\frac{y^2}{2\sigma^2}\right)
\left(1+\sqrt{2\pi}ay\Phi\left(ay\right)\exp\left(\frac{a^2y^2}{2}\right)\right)dy.%
\eeqa
%%%%%%%%%%%%%%%%%%%!!!!!AÑADIDO ANGEL
Note that $\overline{P}_j^{(1)}$ is the probability of error of a coherent binary signaling over a Rayleigh fading channel \citep{proakis01}. The above integral can be solved and expressed in a closed form by using integration by parts, and the result is given by \beqa
\overline{P}_j^{(1)} &=&\frac{1}{2}\left(1-\sqrt{\frac{\frac{\gamma}{2}}{\frac{\gamma}{2}+1}}\right).\label{45}
\eeqa

%The probability of error of for nodes 2 and following is discussed in the Section \ref{sec:imple}

\section{Detectors in the groups of nodes beyond the first group}
\label{sec:imple}

The detectors for the nodes in groups $k>1$ are challenging to implement because of the need to know the joint probabilities of correct decisions of the relaying nodes (see \eqref{eq:dec_rul}).
It is interesting to point out that there is one exception to the need for pmfs, and it is the case of mesh networks with two nodes per group, known CSI, and assumptions as in Claim 2.
%A challenge in implementing the decision rule
%\eqref{eq:dec_rul} at a node in a general scenario is the need for knowing the pmfs of decisions of the nodes   relaying to it, $P({\bf x}^{(k-1)}|x=1)$, $k>2$. %In a simulation environment it is possible to jointly estimate this probability for all the nodes of a given group using Monte Carlo, as we will comment in Section \ref{sec:sim}.
Here, we describe four types of detectors %
for the nodes $R_i^{(k)}$, $k>1$, $i=1, 2,\cdots, n^{(k)}$. We also briefly describe the detectors in multihop networks.

\subsection{Estimation of the joint pmf based on Monte Carlo sampling (MCS)}
\label{sec:est_1}

%Recall that the decision of $R_j^{(k)}$ is based on \eqref{express}.
The joint pmf of interest is $P\left({\bf x}^{(k)}|x\right)$, and we can show that it can be expressed as
\beqa
P\left({\bf x}^{(k)}|x\right)&=&P\left({\rm sgn}\left( \log L_1^{(k)}\right), \;{\rm sgn}\left(\log L_2^{(k)}\right),\; \cdots, \;{\rm sgn}\left(\log L_{n^{(k)}}^{(k)}\right) |x\right),
\eeqa
%In other words, if for example $x_1^{(1)}=1$, $x_1^{(1)}=-1$, $x_3^{(1)}=-1$, and so on, we look for the joint probability of  $P\left(L_1^{(k)}>1, L_2^{(k)}<1, L_3^{(k)}<1, \cdots |x\right)$.
where the random variables $L_i^{(k)}$ are {\em not} independent.

From \eqref{express}, we see that the $L_j^{(k)}$s are expressed via the decisions $\bfx^{(k-1)}$, which allows us to propose a computational method for obtaining the joint pmf. The procedure is the following:
\begin{enumerate}
\item Each node $R_j^{(k)}$ generates samples ${y}_{ij}^{(k)^{(m)}},$ where $i=1, 2, \cdots, n^{(k-1)}$, $m=1, 2, \cdots, M$ from the mixture distribution given by \beqa
    f(y_{1j}, y_{2j}, \cdots,  y_{ij}, \cdots, y_{n^{(k-1)}j})&=&\sum_{{\bf x}^{(k-1)}} P\left({\bf x}^{(k-1)}|x\right)\prod_{i=1}^{n^{(k-1)}} f\left(y_{ij}^{(k)}|x_{i}^{(k-1)}\right).
    \eeqa
\item The node computes the signs of the obtained $\log L_j^{(k)}$, $i=1, 2, \cdots, n^{(k)}$ at the drawn ${y}_{ij}^{(k)^{(m)}}$, $j=1, 2, \cdots, n^{(k)}$ and stores them as a vector in a counter.
\item The process is repeated $M$ times.
\item From the obtained outcomes, the node obtains the estimates of the joint probabilities.
\end{enumerate}

The generation of samples in  the case of a mixture Gaussian is easy. When the channel statistics are known only, the sampling of $y_{ij}$ must come from a distribution given by \eqref{eq:pdf}. To that end, one could apply rejection sampling.

Clearly, at the end, all the nodes in the $k$th group will have different estimates of the joint pmf. The next group may combine all these estimates, for example, by taking the average of the estimates. This approach becomes tedious when the number of nodes in a group gets large. For example, when $n^{(k)}=10$, one has to work with a joint pmf of a 10-dimensional vector, which, in general, requires the computation of 1023 estimates. %For this reason, we have turned our attention to developing alternative ways of estimating the joint pmfs, and they are described in the sequel.
However, there are two ways to decrease the computational load of this approach and they are (1) to have a designated node in a group compute the joint pmf and (2) under an assumption of symmetry, instead of computing the values of $2^{n^{(k)}}-1$ elements, a node computes only $n^{(k)}+1$ elements.

\subsection{Estimation of the joint pmf based on pilot signals (PS)}
\label{sec:est_}

Suppose that  node $R_j^{(k)}$, $k>2$ receives messages from $n^{(k-1)}$ relay nodes, $y_{ij}^{(k)}$. In the rest of the subsection, we suppress the subscript $j$ and the superscripts ${(k)}$. For example, for the received signals $y_{ij}^{(k)}$, we write $y_i$.

Define the variable $z_{i}=1$ if $y_{i}>0$ and $z_{i}=0$ if
$y_{i}<0$. Let also \beqa
\kappa(z_1, z_2, \cdots, z_{n})&=&\sum_{i=1}^{n} z_i 2^{i-1}=\sum_{i=1}^{n} \kappa_i 2^{i-1},
\eeqa
where $\kappa_i\in\{0,1\}$ identifies the $i$th binary digit of
  $\kappa$ represented in binary notation.
Thus, for example, $\kappa=0$ corresponds to $\{y_1<0, y_2<0, \cdots,
y_{n}<0\}$ and  $\kappa=1$ to $\{y_1>0, y_2<0, \cdots, y_{n}<0\}$. In
other words, the joint pmf of $z_1, z_2, \cdots, z_{n}$ can succinctly
be represented by the pmf of $\kappa$ .

%
%
%
%Let the joint pmf of the $z_{i}$ be a vector ${\bf p}_{z}$ with $2^{n^{(k-1)}}$ elements, where for the first element we have
%\beqa
%p_{z,0}&=&P(y_{1}<0,y_{2}<0, y_{3}<0, \cdots, y_{n^{(k-1)}}<0)
%\eeqa
%the second element is
%\beqa
%p_{z,1}&=&P(y_{1}>0,y_{2}<0, y_{3}<0, \cdots, y_{n^{(k-1)}}<0)
%\eeqa
%the third element is
%\beqa
%p_{z,2}&=&P(y_{1}<0,y_{2}>0, y_{3}<0, \cdots, y_{n^{(k-1)}}<0)
%\eeqa
%and so on.

Similarly, we represent the joint pmf of the transmitted symbols of
the relaying nodes. Let $\zeta$ and $\zeta_i$ be defined by
\beq
\zeta(x_1, x_2, \cdots, x_{n})=\sum_{i=1}^{n} {(x_i+1)}
\;2^{i-2}=\sum_{i=1}^{n} \zeta_i2^{i-1},\ \zeta_i\in\{0,1\},
\eeq

and the joint pmf of $x_i$ is equivalent to the pmf of $\zeta$.

We want to estimate $p(\zeta)$ from pilot data (where the source, say,
repeatedly transmits $x=1$). Then the node first estimates the pmf of
$\kappa$ and, based on the model for the channels, estimates the pmf
of $\zeta$. We can write for $\kappa=0, 1, \cdots, 2^n-1$ %
%prefiero usar l para dejar i para los nodos
\beqa
p(\kappa)&=&\sum_{l=0}^{2^n-1} p(\kappa|\zeta=l)
p(\zeta=l),
\eeqa
where the probabilities $p(\kappa|\zeta=l)$ are known, i.e., we can
show that %
\beqa
\label{prob}
p(\kappa|\zeta=l)&=&\prod_{i=1}^{n}p_{c,i}^{\delta_{i}(\kappa,\zeta)}
p_{e,i}^{1-\delta_{i}(\kappa,\zeta)},
\eeqa
where
\beqa
\delta_{i}(\kappa,\zeta)=\left\{\begin{array}{ll}
1,& \kappa_i=\zeta_i\\
0,&{\rm otherwise}
\end{array}
\right.,
\eeqa
where  $\kappa_i=\zeta_i$ means
\beqa
z_i&=&\frac{x_i+1}{2}\nn,
\eeqa
and $p_{c,i}$ is the probability of correct decision based on the
transmitted signal from the $i$th node, and $p_{e,i}$ is the
corresponding probability of error, that is $p_{e,i}=1-p_{c,i}$. This
probability of error is obtained from \eqref{45}.

We can represent the probabilities of $\kappa$ by $${\bf p}_\kappa=\big[p(\kappa=0)\; p(\kappa=1)\; \cdots\; p(\kappa=2^n-1)\big]^\top$$
and the probabilities of $\zeta$ by
$${\bf p}_\zeta=\big[p(\zeta=0)\; p(\zeta=1)\; \cdots\; p(\zeta=2^n-1)\big]^\top.$$
We can also construct a $2^n\times 2^n$ matrix, ${\bf P}$ with probabilities as the ones defined by \eqref{prob}. Then we can write
\beqa
\label{eq:ec_mat}
{\bf p}_\kappa&=&{\bf P}\, {\bf p}_\zeta.
\eeqa
We note that the matrix $\mathbf{P}$  is a doubly stochastic matrix, that is, its rows and columns sum up to one.

We reiterate that we estimate the vector ${\bf p}_\kappa$ from the pilot data, and we denote it by $\widehat{\bf p}_\kappa$. Then  the estimate of ${\bf p}_\zeta$ is readily obtained from
\beqa
\label{solution_2}
\widehat{\bf p}_\zeta&=&{\bf P}^{-1}\, \widehat{\bf p}_\kappa.
\eeqa
This estimate, however, may produce negative probabilities (although the sum of all the elements of $\widehat{\mathbf{p}}_\zeta$ equals one), which is due to errors in estimating ${\mathbf{p}}_\kappa$.
%In the estimation of $\hat{\mathbf{p}}_\lambda$ are substituted for the probabilities of the binary variables $\mathbf{p}_z^{(k)}$ the equality \eqref{eq:ec_mat} is only approximate and the vector solution $\mathbf{p}^{(k)}$ can have negative components.
Thus, we change the set of equations \eqref{eq:ec_mat} to the
following minimization problem with linear constraints:
\begin{equation}
  \label{eq:ec_mod}
  \widehat{\mathbf{p}}_{\kappa}=\mathbf{P} (\mathbf{p}_\zeta+\mathbf{\epsilon}),
\end{equation}
where $\mathbf{\epsilon}$ has minimum norm and adds to 0. In Appendix
2, we show that the solution vector $\widehat{\mathbf p}_\zeta$ to the
system \eqref{eq:ec_mod} has elements given by
\begin{equation}
  \label{eq:sol_cnt}
  [\widehat{\bf p}_\zeta]_i=( b_i-\xi)^+\, ,
%  \begin{cases}
 %   0, & \text{if } b_i \le \xi\\
 %   b_i-\xi, & \text{otherwise}
 % \end{cases},
\end{equation}
where $(x)^+$ is the positive part of $x$, $i=1, 2, \cdots,
2^n$, $\mathbf{b}=\mathbf{P}^{-1} \widehat{\mathbf{p}}_\kappa$, and
$\xi$ is a parameter related to the Lagrange multipliers.
  It is worth pointing out that \eqref{eq:sol_cnt} is in fact an upside-down waterfilling scheme.

In summary, according to this scheme the source node transmits symbols $x$ known to the rest of the network. The nodes in groups $k>1$ can estimate from the received signals from their relaying groups the joint pmfs $P({\bf x}^{(k-1)}|x)$ as just shown. When the estimation is completed, the nodes have the necessary information for processing of the signals with unknown transmitted symbols. % when the transmitted symbols $x$ are unknown. %are able to detect and transmit their decision to nodes of the next group where the procedure is repeated.
Obviously, a drawback of the scheme is that some transmitted symbols are used for estimating joint pmfs and some power is used for estimating the pmfs. % It is interesting to point out that one exception to the need for pmfs, and it is the case of mesh networks with two nodes per group, known CSI, and assumptions as in Claim 2. %We do not discuss this setup further and instead proceed with networks where the nodes need the joint pmfs of correct/incorrect decisions of their relaying nodes.

% pilots are used for estimating the probability of correct decision. However, there is not need for updating the following nodes on the probability of correct decision.

\subsection{Estimation of the joint pmf based on a predefined joint
  pmf (PJP)}

The biggest values of the joint pmf
$P\big(x_1^{(k-1)},x_2^{(k-1)},\cdots,x_{n^{(k-1)}}^{(k-1)}|x=1\big)$ are for combinations when most of the arguments $x_i^{(k-1)}=1$. %$x_1^{(k)}=1,x_2^{(k)}=1,\cdots,x_{n^{(k)}}^{(k)}=1$.
This suggests that we approximate the joint pmf by %a rough approximation of the probability of correct decision:
\begin{equation}
  \label{eq:dep_1}
  P\big(x_1^{(k-1)},x_2^{(k-1)},\cdots,x_{n^{(k-1)}}^{(k-1)}|x=1\big)=
  \begin{cases}
  p_c, & \big|\big\{i:x_i^{(k-1)}=1, i\in\{1, 2, \cdots, n^{(k-1)}\}  \big\}\big|\geq N_f\\
  0, & \text{otherwise}
  \end{cases},
\end{equation}
where $p_c$ is probability obtained from the number of combinations of the elements of ${\bf x}^{(k-1)}$ that satisfy the condition in \eqref{eq:dep_1},  $|\{\cdot\}|$ is the cardinality of the set, and $N_f\leq n^{(k-1)} $. In other words, the probability of correct decision is constant and different from zero whenever the number of arguments $x_i^{(k-1)}=1$ is big enough.  This detector is efficient because only a few terms in the numerator of the likelihood ratio \eqref{express} are different from 0. Claim 2 is used for selecting the significant terms of the denominator of \eqref{express}.  This simple detector does not require the use of pilot signals. % it is not necessary to update the probability of correct decision and could be used in combination with the detector that was already described.
%\bigskip

\subsection{Estimation of the joint pmf based on the assumption of
  independent decisions (ID)}
%\label{sec:idepen}

When for $k>2$ we approximate the joint pmf of ${\bf x}^{(k-1)}$ by \eqref{eq:assumption},
% variables $y_{ij}^{(k)}, i=1,2,\cdots,n^{(k-1)}$ are conditionally independent
for decision making we can use the likelihood ratio \eqref{genexp}. % is used for detecting.
If we take the logarithm on both sides of \eqref{genexp}, and we define
%%%%%%%%%OJO cambiado en el termino negativo P + \tilde L \bar P
\beqa
l_{ij}^{(k)}&=&\log L_{ij}^{(k)}%\nn\\&=&
=\log\left(\tilde{L}_{ij}^{(k)}P_i^{(k-1)}+\overline{P}_i^{(k-1)}\right)-\log\left(P_i^{(k-1)}
+\tilde{L}_{ij}^{(k)}\;\overline{P}_i^{(k-1)}\right),
\eeqa
where
\beqa
\label{14}
l_{ij}^{(k)}&=&\log f({y}_{ij}^{(k)}|x=1) - \log f({y}_{ij}^{(k)}|x=-1),
\eeqa
and
\beqa
\label{eq:lr}
\tilde{L}_{ij}^{(k)}&=&\frac{f({y}_{ij}^{(k)}|x_i^{(k-1)}=1)}{f({y}_{ij}^{(k)}|x_i^{(k-1)}=-1)}
\eeqa
representing the likelihood ratio for the transmitted symbol of $R_i^{(k-1)}$, we can write for the overall loglikelihood ratio
\beqa
\label{rule}
l_{j}^{(k)}&=&\sum_{i=1}^{n^{(k-1)}} l_{ij}^{(k)}.
\eeqa
The decision rule then simplifies to
\beqa
\label{dec_rule}
x_{j}^{(k)}&=&\rm{sgn}\left(l_{j}^{(k)}\right).
\eeqa
%\beq
%\label{dec_rule}
%\begin{array}{lll}
%{\rm If\;} l_{j}^{(k)}&\geq 0,& {\;\rm choose\;} x_j^{(k)}=1\\
%{\rm If\;} l_{j}^{(k)}&< 0,& {\;\rm choose\;} x_j^{(k)}=-1
%\end{array}.
%\eeq

For the probability of error at $R_j^{(k)}$, we can write
%%%OJO cambiado en el termino negativo P + \tilde L \bar P
\beqa
\overline{P}_j^{(k)}&=&P\left(l_{j}^{(k)}<0\big|x=1\right)%\nn\\
=P\left(\sum_{i=1}^{n^{(k-1)}} l_{ij}^{(k)}<0\big|x=1\right)\nn\\
&=&P\left(\sum_{i=1}^{n^{(k-1)}}\left[\log\left(\tilde{L}_{ij}^{(k)}P_i^{(k-1)}+\overline{P}_i^{(k-1)}\right)-
\log\left(P_i^{(k-1)}
+\tilde{L}_{ij}^{(k)}\;\overline{P}_i^{(k-1)}\right)\right]<0\big|x=1\right).\nn\\\label{Pe}
\eeqa
We note that under these assumptions %that the $y_{ij}^{(k)}$s are independent, so must be
the new random variables $l_{ij}^{(k)}$ are independent. Clearly, if the ${y}_{ij}^{(k)}$s are identically distributed, so are the
loglikelihood ratios $l_{ij}^{(k)}$.

The implementation of this detector requires information about the
probability of correct decision of the nodes in the previous group
$P_i^{(k-1)}$, $i=1,\cdots, n^{(k-1)}$. These probabilities can readily be computed as described in subsections A and B. The computations  in this case are much simpler because the problem is broken into $n^{(k-1)}$ independent problems, where in each problem we estimate the probability of detection (error) of only one relaying node at a time. %Clearly, this scheme requires transmission of pilot data.

Another possibility is that the nodes compute their own probabilities of errors from \eqref{Pe} and transmit them to the nodes of the next group. The advantage of this approach is that there is no need for transmitting pilot signals.
%Under the independence assumption a node can evaluate numerically this probability from \eqref{Pe}.
%The probability of correct decision is obtained through sums of random variables, therefore FFT is an efficient method for evaluating the corresponding pdf. After calculating $P_i^{(k)}$ a node must transmit it to nodes in next group.
We note that in the transmission of the probability, some form of quantization must be adopted. % to reduce the amount of transmitted information.
In Section \ref{sec:sim}, we study the influence of quantization  via simulations. %It is important to note that the probability of correct decision $P_i^{(k)}$ calculated under the independence assumption is not equal to the actual value, in fact, it is over-optimistic.

We now briefly describe the implementation of the scheme in the  cases of completely known channels and known channel statistics, respectively.

%With this detector the nodes must also have information about the channels reaching them, as is the case in the detectors described above.

\subsubsection{Completely known channels}
\label{sec:sec_chan_know}

For the conditional distributions of the observations, we can write
\beq
\begin{array}{lll}
f(y_{ij}^{(k)}|x=1) &=& P_i^{(k-1)}{\cal N}(h_{ij}^{(k)},\sigma^2)+\overline{P}_i^{(k-1)}{\cal N}(-h_{ij}^{(k)},\sigma^2)\\
f(y_{ij}^{(k)}|x=-1) &=& \overline{P}_i^{(k-1)}{\cal N}(h_{ij}^{(k)},\sigma^2)+{P}_i^{(k-1)}{\cal N}(-h_{ij}^{(k)},\sigma^2)
\end{array}
\eeq
and for the likelihood ratios,
\beqa
\tilde{L}_{ij}^{(k)}&=&\exp\left(\frac{2y_{ij}^{(k)}h_{ij}^{(k)}}{\sigma^2}\right),
\label{likrat}
\eeqa
whereas for the loglikelihood terms we have
%%%%%OJO Cambiado el exp positivo multiplica a \bar P
\beqa
l_{ij}^{(k)}&=&
\log\left(\exp\left({\frac{2y_{ij}^{(k)}h_{ij}^{(k)}}{\sigma^2}}\right)P_i^{(k-1)}+\overline{P}_i^{(k-1)}\right)\nonumber\\
&-&\log\left({P}_i^{(k-1)}+\exp\left({\frac{2y_{ij}^{(k)}h_{ij}^{(k)}}{\sigma^2}}\right)\overline{P}_i^{(k-1)}\right).
\label{eq:genexpr}
\eeqa
We can also find the distribution of $l_{ij}^{(k)}$ by change of variables. We obtain
%%OJO cambiado el \pi en la raiz
\begin{equation}
  \label{eq:sim1}
  \begin{array}{lll}
    f(l_{ij}^{(k)}|x=1)& = & \frac{1}{2\sqrt{ 2 \pi}}\frac{\sigma}{h_{ij}^{(k)}} \frac{e^{l_{ij}^{(k)}}(2P_i^{(k-1)}-1)}
           {\left(\overline{P}_i^{(k-1)}-P_i^{(k-1)} e^{l_{ij}^{(k)}} \right)
            \left(\overline{P}_i^{(k-1)} e^{l_{ij}^{(k)}} - P_i^{(k-1)}\right)}\\
          & \times& \bigg( P_i^{(k-1)} e^{-\frac{1}{2}\big(\frac{\sigma}{h_{ij}^{(k)}}\frac{1}{2}
              \log \frac{\overline{P}_i^{(k-1)}-P_i^{(k-1)} e^{l_{ij}^{(k)}} }
                   {\overline{P}_i^{(k-1)} e^{l_{ij}^{(k)}} - P_i^{(k-1)} }-\frac{h_{ij}^{(k)}}{\sigma}\big)^2 }\\
          & + &     \overline{P}_i^{(k-1)} e^{-\frac{1}{2}\big(\frac{\sigma}{h_{ij}^{(k)}}\frac{1}{2}
              \log \frac{\overline{P}_i^{(k-1)}-P_i^{(k-1)} e^{l_{ij}^{(k)}} }
                   {\overline{P}_i^{(k-1)} e^{l_{ij}^{(k)}} - P_i^{(k-1)} }+\frac{h_{ij}^{(k)}}{\sigma}\big)^2 } \bigg)
  \end{array}.
\end{equation}

The probabilities of error  of the nodes in group $k\geq 2$ are obtained by the recursive equation \eqref{Pe}. A closed form analytical solution of the recursive equation in the general case, however, cannot be obtained. % except in some special cases.

\subsubsection{Channels with known statistics}
\label{sec:sec_knst}

%Under the independence assumption
The probability of error at the nodes in group $k\geq 2$ is the same, that is, it is given by \eqref{Pe}, where now
%\beqa
%\overline{P}_j^{(k)}&=&P\left(l_{j}^{(k)}<0\big|x=1\right)\nn\\
%&=&P\left(\sum_{i=1}^{n^{(k-1)}} l_{ij}^{(k)}<0\big|x=1\right)\nn\\
%&=&P\left(\sum_{i=1}^{n^{(k-1)}}\left[\log\left(\tilde{L}_{ij}^{(k)}P_i^{(k-1)}+\overline{P}_i^{(k-1)}\right)-\log\left(\tilde{L}_{ij}^{-1^{(k)}}P_i^{(k-1)}
%+\overline{P}_i^{(k-1)}\right)\right]<0\big|x=1\right)\nn\\\label{Pe1}
%\eeqa
\beqa
\tilde{L}_{ij}^{(k)}%&=&\frac{f({y}_{ij}^{(k)}|x_{ij}^{(k-1)}=1)}{f({y}_{ij}^{(k)}|x_{ij}^{(k-1)}=-1)}\nn\\
&=&\frac{1+\sqrt{2\pi}a y_{ij}^{(k)}
\Phi(ay_{ij}^{(k)})\exp\left(\frac{a^2y_{ij}^{(k)^2}}{2}\right)}
{1-\sqrt{2\pi}a y_{ij}^{(k)}\Phi(-ay_{ij}^{(k)})\exp\left(\frac{a^2y_{ij}^{(k)^2}}{2}\right)}
\eeqa
%where
%\beqa
%y_{ij}^{(k)}&\sim& f(y_{ij}^{(k)}|x_{ij}^{(k-1)}=1)\nn.
%\eeqa
is obtained by using \eqref{eq:lr} and \eqref{eq:pdf}.
The probabilities of error  of the nodes in groups $k\geq 2$ are again obtained  numerically by the recursive equation \eqref{Pe}.

\subsection{Detectors for multihop networks}
\label{sec:multh}

In multihop networks, the variables $y_{ij}^{(k)}, i=1,2,\cdots,n^{(k-1)}$, $k\geq 1$, are conditionally independent. Therefore, for detection at the relay nodes one uses the likelihood ratio in \eqref{genexp}. The last node in the network, $R_1^{(K+1)}$, collects the information of all the nodes from the previous group and uses the decision rule \eqref{dec_rule}, where the sum loglikelihood $l_{j}^{(k)}$ is calculated from \eqref{rule}. For this node, the obtained results under the independence assumption apply. % for the two scenarios considered in this paper: completely known channels and knowledge of their statistics.

\section{Simulations}
\label{sec:sim}

We conducted many experiments where we compared the performances of
the mesh and multihop networks shown in Figs. \ref{fig:mesh} and
\ref{fig:mhop}, respectively. In the experiments, the channels were
Rayleigh distributed and, for better understanding of the results and
easier comparisons, the SNRs were the same for all the nodes. Most of
the simulations were for the scenario where the channel statistics are
known. This case presents two main advantages in real networks:
  (a) the information about the statistics is provided just once for
  each node and (b),
  the probabilities of correct detection are evaluated also once. By
  contrast, when detecting with complete CSI at the nodes, the
  channel information must be updated if the CSI changes, and the
  probability of correct detection has to be evaluated accordingly.

 %However, a comparison has been done between mesh and multihop networks with Rayleigh channels known by the nodes.}

\begin{figure}[t]
  \centerline
  {
    \includegraphics[width=12cm,angle=270]{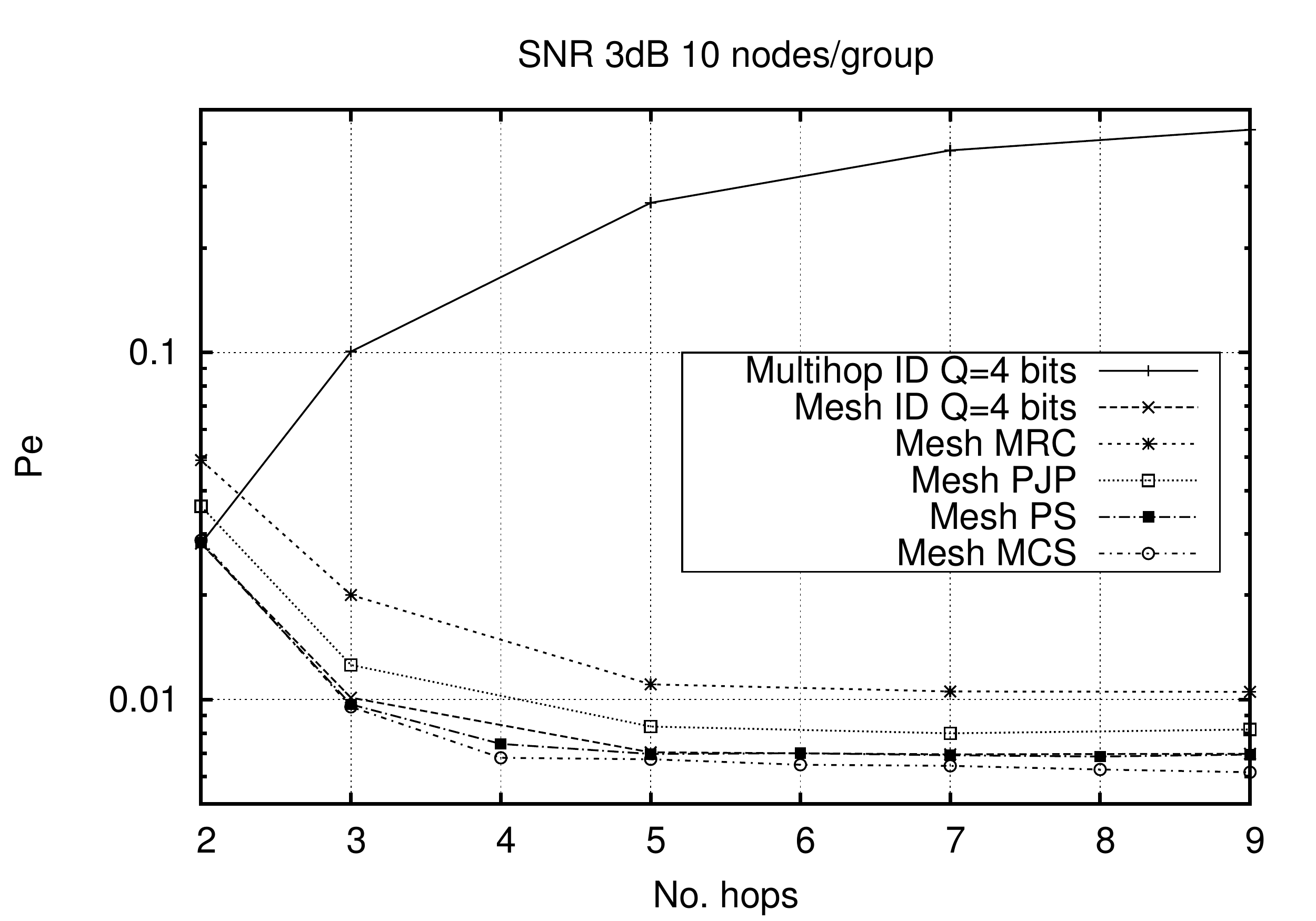}
  }
  % \vspace*{-0.4cm}
  \caption{Probabilities of error for mesh and multihop networks as functions of number hops for the case of known statistics.% and an SNR of 3dB. The descriptions of the detectors used by the nodes are described in the text.
  %Simulation results of mesh networks with detectors that
   % consider the dependence among the decisions of the nodes and
    %suboptimal detectors based on the independence assumption. For
    %comparison the results obtained with the optimal quantized
    %detector for multihop networks are shown and the ones obtained
    %with the MRC detector.
    }
  % \vspace*{0.5cm}
  \label{fig:dep1}
\end{figure}

First, we considered a mesh network with a decision rule based on
complete knowledge of the probabilities of correct decision of the
nodes of the previous groups and based on \eqref{eq:dec_rul}. The
results of this setup were a benchmark for comparison. The joint pmfs
of correct/incorrect decisions of the groups for $k>1$ were computed
using the MCS method.

In Fig. \ref{fig:dep1}, we show in the curve Mesh MCS the probabilities of error of mesh networks with different number of hops when all the joint pmfs were known. % is shown in  Fig. \ref{fig:dep1}. %, we show the results of error probability versus the number of hops of a mesh network where all the nodes in a group have access to the probabilities of correct decision of the previous group. The dependence among the decisions of the nodes in a given group is taken into account in a complete way.
In the simulations, the number of nodes per group was 10 and the SNR $=$ 3 dB. The error probability decreased as the number of hops increased until a point where it remained constant with the number of hops. The spatial redundancy introduced by the mesh network lowered the error probability until an error floor was reached.

In Fig. \ref{fig:dep1}, we also present the performance of the mesh
network using the PS method. % when the joint pmf of $P(\mathbf{x}_{i}^{(k)}|x=1)$ was  estimated from pilot data (the curve is labeled with DEPEN-EST). %, we present the results of error probability obtained estimating the probability of correct decision, using the same conditions already commented. The error probability is obtained using Monte Carlo and following the procedure described in Section \ref{sec:est_}.
As can be seen, the obtained results are close to the ones of mesh networks that have complete information about the pmfs.

To these curves, four other curves are also displayed. One of them shows the performance of %The results obtained with
the  PJP detector. % from subsection C of Section V (curve labeled DEPEND-SIMPLE). % of Fig. \ref{fig:dep1}. Again Monte Carlos simulations have been used for obtaining the error probability and the conditions are the same.
With the curve Mesh ID Q=4 bits, we show the performance of the
detector based on the ID detector, % independence assumption,
and where the probabilities of correct decision are quantized with
four bits. Note that a node uses the probabilities of correct
  decision of previous nodes and  quantizing these variables reduces the
  amount of information to be distributed.
%ing the probability of correct decision of the previous node.
%gives results that are very close to the optimum case with complete information about the probability of correct decision. In Fig.  \ref{fig:dep1}, curve MESH Q-4 bits, we show the results obtained when 4 bits are used for quantizing the probability of correct decision of the previous node.
Finally, with the Multihop ID Q=4 bits %MULTIHOP Q-4
curve we plotted the performance of the multihop network, and with the MRC curve, we displayed the performance of the MRC detector \citep{proakis01}. For the multihop network, the probabilities of correct decision of the previous node were again quantized with four bits. The MRC detector was implemented by combining the signals arriving to a node using the MRC criterion.

The worst performance was clearly achieved by the multihop network. Also, its performance deteriorated steadily as the number of hops increased. The performance of the network whose nodes had to estimate the joint decision pmfs of the nodes in the previous group performed almost as well as the network whose nodes knew the joint pmfs. Surprisingly well performed the mesh network that used the independence assumption and where the nodes quantized their probabilities  with four bits. Next in performance of the mesh networks came the one that employed the simple detector and finally, the worst was the network with MRC detectors.

The analytical study of the quantization effect on the ID
  detectors is difficult. %, if possible, but it is affordable to extract conclusions from simulations. In fact, in Algorithm \ref{alg:mtber} the quantizing and updating to the next node has been included as one of the steps of the algorithm.
Instead, we performed experiments where the number of bits for
representing the quantized probabilities was either one bit (the
quantized probabilities were zero and one) or four bits (for 16
probability values, that is, the quantized probabilities were 0,
0.0666, $\cdots$, 1). We also obtained results when the probabilities
were not quantized. The results are shown in
Fig. \ref{fig:idep} . There we see the probability of error as a
function of the number of hops. The nodes assumed independence of the
decisions of the nodes of the previous group. The SNR was 3 dB. % and the number of nodes per group was 5 or 10. The curves with 5 and 10 nodes per group are labeled with 5 n/g and 10n/g, respectively. The labels also contain information about the used number of bits. % used for quantizing.
%In the curve with infinity bits no quantizing has been done.
In all the mesh networks with 10 nodes per group, the BER decreased with the number of hops. %This effect has been commented before.
The results show that four bits were enough for achieving almost as good performance as the one when there was no quantization. %to describe the probability of correct decision.
Interestingly, using only one bit for quantization also yielded good results. %this probability gives results that are close to the no quantizing case.
Note that representing the probability of correct decision with one bit is a form of selection combining: the nodes with low probability withdraw themselves from combining in the following group of nodes.

In the case of five nodes per group (curve Mesh ID 5 nods/grp
  Q=4 bits) the decrease in probability of error probability was slower. The performance of this network was worse than that with 10 nodes per group by an order of magnitude.
In Fig. \ref{fig:idep}, we also plotted the performance of two multihop networks. One of them was the same network whose performance is shown in Fig. \ref{fig:dep1}, and the other  corresponds to the multihop network with CMRC detectors (Multihop CMRC 10 nods/grp) \citep{wang07}.
%, we present the results obtained with the CMRC
%detector for multihop networks described in \citep{wang07}.
%%%OJO MODIFICADO ESTE PARRAFO
Note that the CMRC detector uses complete information about the
channel.  Nevertheless, it has a larger error probability than the
multihop ID detector (curve
Multihop ID 10 nods/grp Q=4 bits), where only channel statistics are used. In both cases the error probability increases very fast with the number of hops.  %However, as the number of hops increases, the  error of the network with CMRC detectors  increases faster with the number of hops than the error of the other network.

%We have shown the in Fig. \ref{fig:dep1} that with multihop networks the error probability increases with the number of hops. We have
%reproduced in Fig. \ref{fig:idep}, curve MULTIHOP 10n/g 4 bits, that results to make an easier comparison. In the same figure, curve
%MULTIHOP CMRC 10n/g, we present the results obtained with the CMRC detector for multihop networks described in \citep{wang07}. This detector uses complete information about the channel and for this reason it has lower error probability that the multihop detector in the curve MULTIHOP 10n/g 4 bits, where only channel statistics are used. However, as the number of hops increases, the CMRC's error probability increases.

{%\singlespacing
%\begin{algorithm}[htb]
%  %\SetLine
%  \KwData{Topology of network, channel and noise statistics}
%  \KwResult{BER}
%  \Begin{
%    set all channels statistics and noise variances\;
%    set a list with nodes' processing order\;
%    \For{error count $\leq$ fixed maximum number of errors}{
%      send a random bit $x_1^{(0)}$ in the fist node $R_1^{(0)}$\;
%      generate the channels from a  Rayleigh generator\;
%      generate the noise from a Gaussian source\;
%      \ForEach{node $R_j^{(k)}$}{
%        obtain the received samples $y_{ij}^{(k)}$;
%        pick the probabilities of correct decision $P_{i}^{(k-1)}$\;
%        calculate the loglikelihood $l_j^{(k)}$\;
%        estimate and send $x_j^{(k)}$\;
%        \uIf(once in the simulation ){not yet calculated}{
%          calculate $P_{i}^{(k)}$\;
%          \lIf{have to quantize} {quantize $P_{i}^{(k)}$}\;
%          send $P_{i}^{(k)}$\;
%        }
%      }
%      received bit $=x_1^{(K+1)}$\;
%      \lIf{ $x_1^{(0)}\neq x_1^{(K+1)}$}{increase error count by one}\;
%      increase run count by one\;
%    }
%    calculate the BER\;
%  }
%  \label{alg:mtber}
%  \caption{Simulation under independence among the decisions of the nodes}
%\end{algorithm}
%}

\begin{figure}[thb]
  \centerline
  {
    \includegraphics[width=12cm,angle=270]{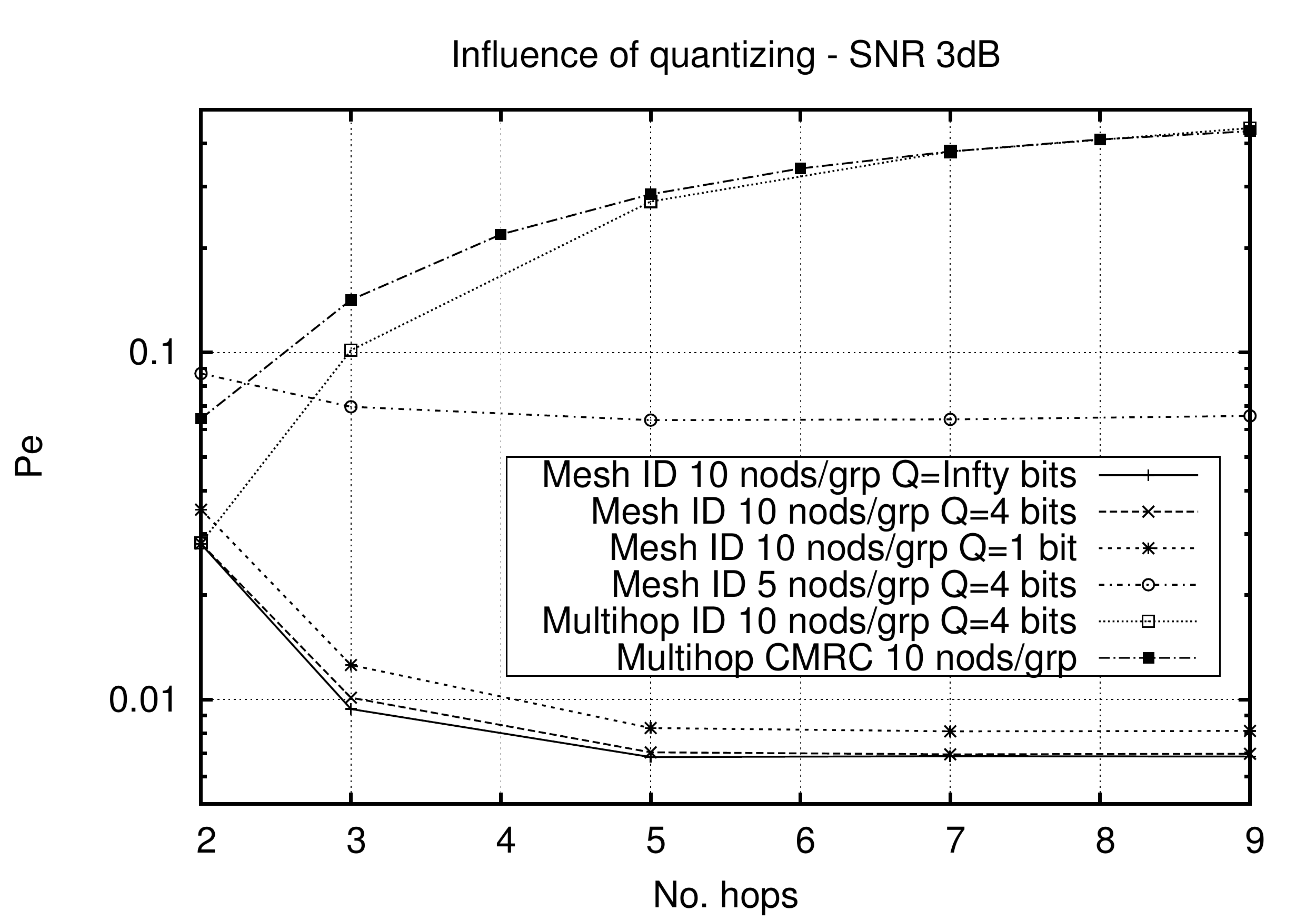}
  }
  % \vspace*{-0.4cm}
  \caption{Probabilities of error for mesh and multihop networks as functions of number hops for MAP detectors that used conditional
    independence among the probabilities of correct decision. The nodes knew the statistics of the channels and the SNR was 3dB.}
%  Simulation results obtained in the mesh and multihop
 %   networks using the MAP decision rule and assuming conditional
  %  independence between the probabilities of correct decision.}
   \vspace*{0.5cm}
  \label{fig:idep}
\end{figure}

%We have shown the in Fig. \ref{fig:dep1} that with multihop networks
%the error probability increases with the number of hops. We have%
%reproduced in Fig. \ref{fig:idep}, curve MULTIHOP 10n/g 4 bits, that
%results to make an easier comparison. In the same figure, curve
%MULTIHOP CMRC 10n/g, we present the results obtained with the CMRC
%detector for multihop networks described in \citep{wang07}. This
%detector uses complete information about the channel and for this
%reason it has lower error probability that the multihop detector in
%the curve MULTIHOP 10n/g 4 bits, where only channel statistics are
%used. However, as the number of hops increases, the CMRC's error
%probability increases.

\begin{figure}[thb]
  \centerline
  {
    \includegraphics[width=12cm,angle=270]{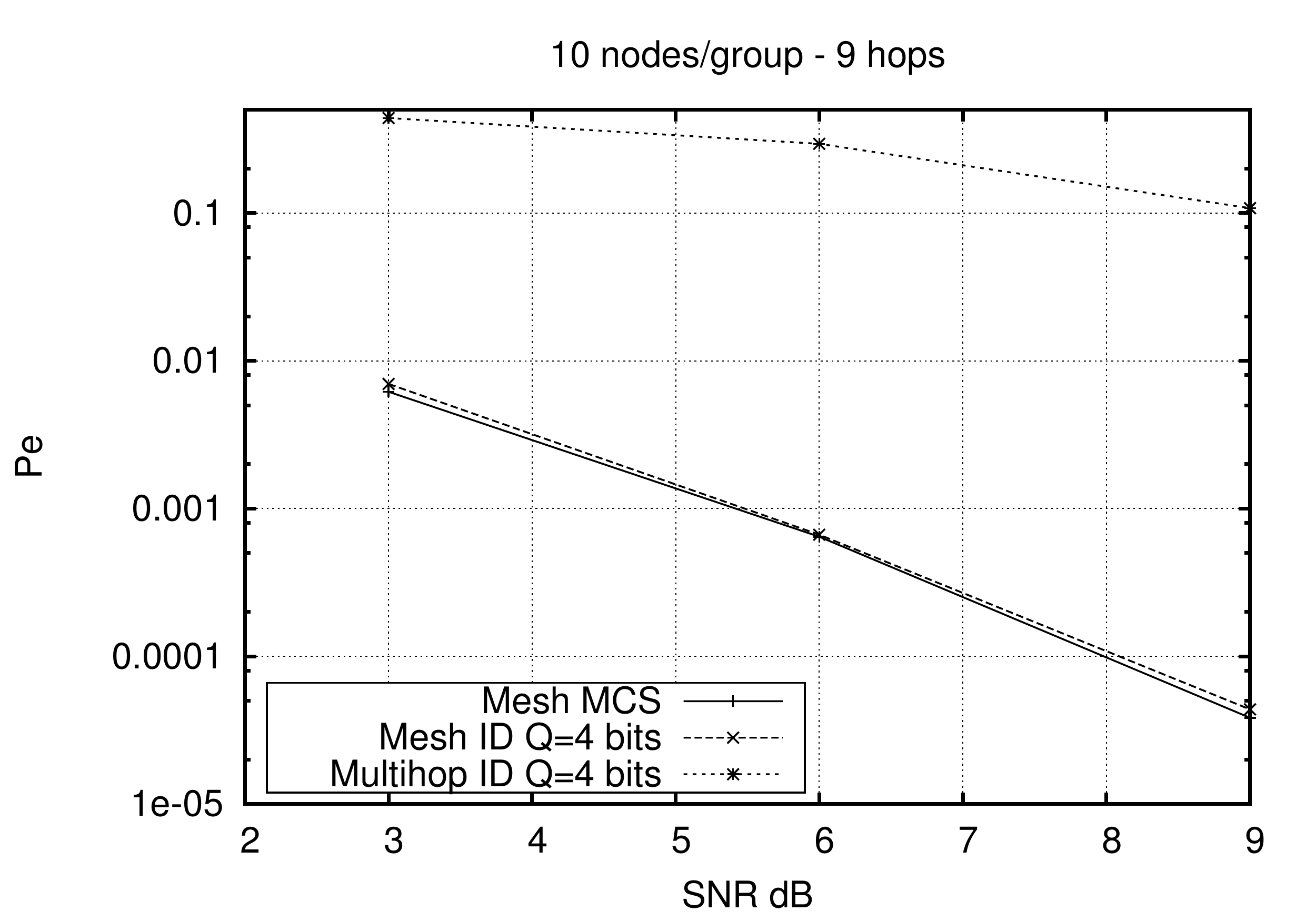}
  }
  % \vspace*{-0.4cm}
  \caption{%Probabilities of error for mesh and multihop networks as
    %functions of SNR for MAP detectors that used conditional and unconditional
    %independence among the probabilities of correct decision.
Probabilities of error for mesh and multihop networks as functions of SNR for MAP detectors.
  %  Simulation results obtained in the mesh and multihop
   % networks using the MAP decision rule and assuming conditional
%    independence between the probabilities of correct decision.
    }
  % \vspace*{0.5cm}
  \label{fig:idepsn}
\end{figure}

Finally, in Fig. \ref{fig:idepsn} we present curves of probability of
error versus SNR for a mesh network with MCS detectors, a mesh
network with ID detectors and four bits for quantization, and a multihop network. The number of nodes per group was 10 and the curves were obtained after 9 hops.
%The cases studied are: simulation with complete knowledge of the probability of correct decision of previous nodes in a mesh network, simulation assuming independence in a mesh network and numerical evaluation of the decision rule \eqref{dec_rule} over a multihop network.
From the experiment we can draw the following conclusions: a) the assumption of independence is adequate and gives results that are very close to the case of dependence between the probabilities of correct decision of the nodes in a given group and b) the mesh networks offer great advantages in performance over multihop networks. %, and  c) all the cases have the same diversity order, determined by the number of nodes per group of the last hop.

\section{Conclusions}
\label{sec:con}

In this paper we studied the performance of mesh networks whose nodes
operate in the decode-and-forward mode. In particular, we
investigated the probability of error in these networks as a function
of SNR, number of nodes per group, and number of hops. We studied  two
cases, where in the first case the nodes knew the CSI and in the
second, the nodes worked only with the statistics of the channels. We
compared the mesh networks with multihop networks and showed the gain
in performance of the former with respect to the latter. In the presented work we dealt with binary modulations, but the generalization of the proposed detectors to deal with $M-$ary, $M>2$, modulations is not difficult.

\section*{Appendix 1}

%\subsection*{Proof of Claim 1}
%
%For $k=1$ the claim is an obvious consequence of the independence of the channels from $R_1^{(0)}$ to nodes of the first group and the independence of the noise variables of nodes in that group. For $k=2$ we have
%
%\beqa
%f\left({\bf y}_j^{(2)}|x\right)&=&\sum_{{\bf x}^{(1)}}f\left({\bf y}_j^{(2)}|{\bf x}^{(1)}\right)P({\bf x}^{(1)}|x)\\
%&=&\sum_{{\bf x}^{(1)}}%
%\prod_{i=1}^{n^{(1)}} f\left({y}_{ij}^{(2)}|{x}_i^{(1)}\right)P({x}_i^{(1)}|x)\\
%&=& \prod_{i=1}^{n^{(1)}}  \sum_{{x}_i^{(1)}}%
% f\left({y}_{ij}^{(2)}|{x}_i^{(1)}\right)P({x}_i^{(1)}|x)
%\label{tpt}
%\eeqa
%proving the claim.
%\hfill$\Box$

\subsection*{Proof of Claim 2}
We prove the claim by induction. Suppose that the channel
  likelihoods satisfy $f\big(y_{ij}^{(k)}|x_i^{(k-1)}\big)=f\big(-y_{ij}^{(k)}|-x_i^{(k-1)}\big)$
and
$f\big(-y_{ij}^{(k)}|x_i^{(k-1)}\big)=f\big(y_{ij}^{(k)}|-x_i^{(k-1)}\big)$,
and that the claim is true for $k-1$,
i,e. $P(\mathbf{x}^{(k-1)}|x=1)=P(-\mathbf{x}^{(k-1)}|x=-1)$  Then we
have the likelihood ratio \eqref{express}
\begin{align}
  L_j^{(k)}(\mathbf{y}_j^{(k)})
  &=\frac{\sum_{{\bf x}^{(k-1)}}\prod_{i=1}^{n^{(k-1)}}
    f({y}_{ij}^{(k)}|{ x_i}^{(k-1)})P({\bf x}^{(k-1)}|x=1)}
  {\sum_{{\bf x}^{(k-1)}}
    \prod_{i=1}^{n^{(k-1)}}f({y}_{ij}^{(k)}|{x_i}^{(k-1)})
       P({\bf x}^{(k-1)}|x=-1)} \notag\\
&=\frac{\sum_{{\bf x}^{(k-1)}}\prod_{i=1}^{n^{(k-1)}}f({-y}_{ij}^{(k)}|{
    -x_i}^{(k-1)})P({-{\mathbf{x}}}^{(k-1)}|x=-1)}{\sum_{{\bf x}^{(k-1)}}
    \prod_{i=1}^{n^{(k-1)}}f({-y}_{ij}^{(k)}|{
    -x_i}^{(k-1)})P(-{\mathbf{x}}^{(k-1)}|x=1)},
\label{eq:expr_2_1}
\end{align}
where we have used the inductive hypothesis and the symmetry of the likelihoods.
We can reorder the sums by replacing $\mathbf{x}$ with $-\mathbf{x}$ to get
\begin{align}
  L_j^{(k)}(\mathbf{y}_j^{(k)})
  &=\frac{\sum_{{\mathbf{x}}^{(k-1)}}\prod_{i=1}^{n^{(k-1)}}
       f({-y}_{ij}^{(k)}|{x}_i^{(k-1)})
         P({\mathbf{x}}^{(k-1)}|x=-1)}
     {\sum_{{\mathbf{x}}^{(k-1)}}\prod_{i=1}^{n^{(k-1)}}
       f({-y}_{ij}^{(k)}|{x}_i^{(k-1)})
        P({\mathbf{x}}^{(k-1)}|x=1)} \notag\\
&=\frac{1}{L_j^{(k)}(-\mathbf{y}_j^{(k)})}.
\label{eq:expr_2_2}
\end{align}
From  \eqref{eq:expr_2_2}, we have the equality
\begin{multline}
  \label{eq:ec_ps}
  P\big(L_1^{(k)}(\mathbf{y}_1^{(k)})\gtreqless 1,\cdots,
  L_j^{(k)}(\mathbf{y}_j^{(k)})\gtreqless 1, \cdots |x=1\big)=\\
  P\big(L_1^{(k)}(-\mathbf{y}_1^{(k)})\lesseqgtr 1,\cdots,
  L_j^{(k)}(-\mathbf{y}_j^{(k)})\lesseqgtr 1, \cdots |x=1\big).
\end{multline}
Next we show that
\begin{align}
  f(-\mathbf{y}_j^{(k)} |x=1)&=
      \sum_{\mathbf{x}^{(k-1)}}\prod_{i=1}^{n^{(k-1)}}
          f({-y}_{ij}^{(k)}|{ x_i}^{(k-1)})P(\mathbf{x}^{(k-1)}|x=1)\notag\\
    &=\sum_{\mathbf{x}^{(k-1)}}\prod_{i=1}^{n^{(k-1)}}
          f({y}_{ij}^{(k)}|{-x_i}^{(k-1)})P(-{\mathbf{x}}^{(k-1)}|x=-1)\notag\\
    &=f(\mathbf{y}_j^{(k)} |x=-1)
\label{eq:p-y}
\end{align}

and therefore
\begin{multline}
  \label{eq:ec_psf}
  P\big(L_1^{(k)}(\mathbf{y}_1^{(k)})\gtreqless 1,\cdots,
  L_j^{(k)}(\mathbf{y}_j^{(k)})\gtreqless 1, \cdots |x=1\big)=\\
  P\big(L_1^{(k)}(\mathbf{y}_1^{(k)})\lesseqgtr 1,\cdots,
  L_j^{(k)}(\mathbf{y}_j^{(k)})\lesseqgtr 1, \cdots |x=-1\big)
\end{multline}
proving the claim for the $k$th step.

The proof of the  claim for $k=1$ is immediate. % and was commented above in \eqref{eq:sim_1} and \eqref{eq:sim_2}.
This completes the proof of Claim 2.

\section*{Appendix 2}
%\subsection*{Solution}
\label{sec:app}

The system \eqref{eq:ec_mod} can be expressed as the following optimization problem:

\begin{equation}
  \label{eq:ec_can}
  \begin{aligned}
    \text{minimize}\ \  & \sum \epsilon_i^2\\
    \text{subject to}\ \ &p_i^{(k-1)}+\epsilon_i=b_i,\ i=1,\cdots,n^{(k-1)}\\
    &-p_i^{(k-1)}\le 0,\  i=1,\cdots,n^{(k-1)}\\
    &\sum \epsilon_i=0.
  \end{aligned}
\end{equation}
The last two conditions make sure that the vector $\mathbf{p}^{(k-1)}$ is a probability vector because, as can easily be proved, $\mathbf{b}$ is a vector whose elements add to one.
The problem \eqref{eq:ec_can} is convex and for it the Karush-Kuhn-Tucker (KKT) condition  is necessary and sufficient for the optimal solution \citep{luo06}. More specifically, the KKT condition is
\begin{equation}
  \label{eq:ec_kkt}
  \begin{aligned}
      \nabla \;\sum \epsilon_i^2 &+ \sum_{i=1}^{n^{(k-1)}}
      \lambda_i \nabla (-p_i^{(k-1)}-1) +
      \sum_{i=1}^{n^{(k-1)}} \nu_i \nabla (p_i^{(k-1)}+\epsilon_i-b_i) +
      \nu_{e}\nabla \sum \epsilon_i = 0\\
      &-p_i^{(k-1)}\le 0 \quad i=1,\cdots,n^{(k-1)}\\
      &p_i^{(k-1)}+\epsilon_i=b_i\quad i=1,\cdots,n^{(k-1)}\\
      &\sum \epsilon_i = 0\\
      &\lambda_i (-p_i^{(k-1)}-1)=0, \ \ \lambda_i\geq 0, \quad i=1,\cdots,n^{(k-1)}.
  \end{aligned}
\end{equation}
After applying the partial derivatives, for the KKT condition we obtain
\begin{equation}
  \label{eq:ec_kkt_2}
  \begin{aligned}
      \epsilon_i&+\frac{1}{2}(\nu_i+\nu_e)=0; &\quad i=1,\cdots,n^{(k-1)}\\
      &\lambda_i=\nu_i;&\quad i=1,\cdots,n^{(k-1)}\\
      &\lambda_i\geq 0, \lambda_i=0 \ \text{if} \ p_i^{(k-1)}\geq 0,&
      \quad i=1,\cdots,n^{(k-1)}\\
      &p_i^{(k-1)}\geq 0, &\quad i=1,\cdots,n^{(k-1)}\\
      &\sum \epsilon_i = 0 &\\
      &p_i^{(k-1)}+\epsilon_i=b_i&\quad i=1,\cdots,n^{(k-1)}.
  \end{aligned}
\end{equation}
We define the variable
\begin{equation}
  \label{eq:ec_kapp}
  \xi=\frac{\sum \lambda_i}{M},
\end{equation}
where $M$ is the number of variables $p_i^{(k-1)}$ equal to zero. With it and the KKT condition \eqref{eq:ec_kkt_2}, the solution to the system \eqref{eq:ec_can} can be expressed in a compact way as given by \eqref{eq:sol_cnt}.

\bibliographystyle{unsrtnat}
\bibliography{msh_pap_bib}
%\bibliography{IEEEabrv,msh_pap_bib}

\end{document}